\documentclass[useAMS,usenatbib,usegraphicx]{mn2e}
\usepackage{color}
\definecolor{dark-blue}{rgb}{0.15,0.15,0.4}
\usepackage[ocgcolorlinks]{hyperref}
\hypersetup{citecolor=dark-blue, linkcolor=dark-blue}
\usepackage{amsmath}
\usepackage[T1]{fontenc}
\usepackage{aecompl}
\usepackage{amsfonts}
\usepackage{amssymb}
\usepackage{times}

\title[XRB Formation in BCD Galaxies]{X-ray binary formation in low-metallicity blue compact dwarf galaxies}
\author[M. Brorby, P. Kaaret, and A. Prestwich]{M. Brorby$^{1}$\thanks{E-mail:
matthew-brorby@uiowa.edu}, P. Kaaret$^{1}$, and A. Prestwich$^{2}$\\
$^{1}$Department of Physics and Astronomy, University of Iowa, Iowa City, IA 52242\\
$^{2}$Harvard-Smithsonian Center for Astrophysics, 60 Garden street, Cambridge, MA 02138}
\begin{document}

\pagerange{\pageref{firstpage}--\pageref{lastpage}} \pubyear{2014}

\maketitle


\label{firstpage}

\begin{abstract}
X-rays from binaries in small, metal-deficient galaxies may have contributed significantly to the heating and reionization of the early universe. We investigate this claim by studying blue compact dwarfs (BCDs) as local analogues to these early galaxies.
We constrain the relation of the X-ray luminosity function (XLF) to the star-formation rate (SFR) using a Bayesian approach applied to a sample of 25 BCDs. The functional form of the XLF is fixed to that found for near-solar metallicity galaxies and is used to find the probability distribution of the normalisation that relates X-ray luminosity to SFR.
Our results suggest that the XLF normalisation for low metallicity BCDs (12+log(O/H) $<$ 7.7) is not consistent with the XLF normalisation for galaxies with near solar metallicities, at a confidence level $1-5\times 10^{-6}$. The XLF normalisation for the BCDs is found to be $14.5\pm 4.8\ (M_\odot^{-1}\, \text{yr})$, a factor of $9.7\pm 3.2$ higher than for near solar metallicity galaxies. Simultaneous determination of the XLF normalisation and power law index result in estimates of $q = 21.2^{+12.2}_{-8.8}\ (M_\odot^{-1}\, \text{yr})$ and $\alpha = 1.89^{+0.41}_{-0.30}$, respectively. Our results suggest a significant enhancement in the population of high-mass X-ray binaries in BCDs compared to the near-solar metallicity galaxies. This suggests that X-ray binaries could have been a significant source of heating in the early universe.
\end{abstract}


\section{Introduction}\label{intro}
X-rays may have played an important role in heating the early universe \citep{shull1985,haardt1996,mirabel,mcquinn2012,mesinger2013}. However, direct study of the sources that produced X-rays in the early universe is currently impractical due to their high redshifts. \citet{kunth} have suggested that blue compact dwarf (BCD) galaxies may be local analogues of the X-ray producing, metal-deficient galaxies found in the early universe. The X-ray emission of BCDs is dominated by point-like sources that are likely high-mass X-ray binaries (HMXB) \citep{thuan}. Recently \citet{mapelli2010}, \citet{kaaret}, and \citet{prestwich} have reported enhanced production of X-ray binaries in low-metallicity, or extremely metal poor galaxies (XMPG), relative to the star formation rate (SFR).

To further study this enhanced production, we analyse data from a sample of 25 metal-poor BCDs. We compare our results with those of near-solar metallicity galaxies from a study done by \citet{mineo}. Initially, we assume that the X-ray luminosity function (XLF) power law slope is the same for low and near-solar metallicities. Using a Bayesian approach, we show that the distributions of the XLF normalisations for the low-metallicity BCDs are significantly different from that found for near-solar metallicity galaxies. We also find the most probable values for both the power law slope and normalisation of the metal poor XLF and find that the slope is consistent with that derived for near-solar metallicity galaxies.

In Section~\ref{sect:sample}, we discuss the selection criteria for our sample of dwarf galaxies.
Section~\ref{sect:analysis} contains the full description of the procedures we used to prepare and analyse our dataset. Section~\ref{sect:bayes} introduces the Bayesian approach we used in determining the XLF normalisation. In Section~\ref{sect:results}, we present our results along with a short discussion regarding uncertainty in SFRs.

\section{Galaxy Sample}\label{sect:sample}
The analysis was done using a sample of BCDs that is given in Table~\ref{PropTable}. We selected galaxies with metallicity such that $12+\log(\text{O}/\text{H}) < 7.7$ and a distance $<30$~Mpc. The metallicity cut-off was chosen to select galaxies similar to those found at high redshifts~\citep{savaglio2006}. The metallicities in Table~\ref{PropTable} were taken from \citet{izotov2007}.

We selected nearby galaxies so that the individual, compact X-ray sources are bright enough to be detected. We searched the literature for BCDs fitting our criteria, with sources including \citet{kunth}, \citet{hopkins}, \citet{wu}, \citet{gil de paz}, \citet{papaderos}, \citet{izotov2010}, and references therein. We found 27 BCDs, of which 25 had been observed by the \textit{Chandra X-ray Observatory}\footnote{\url{http://cxc.harvard.edu}}. None of the galaxies in our sample were originally discovered in X-rays. We use the 25 BCDs that have been observed by \textit{Chandra} as our sample. The luminosity limits for which the \textit{Chandra} data are complete varies and are listed in Table~\ref{PropTable}. We discuss the method for determining completeness limits in Section~\ref{subsect:completeness}.
\begin{table*}
\centering
\begin{minipage}{120mm}
\caption{BCD Sample}\label{PropTable}
\begin{tabular}{l r c c c c c}
\hline
\hline\noalign{\vspace{1mm}}
Name	&	ObsID	&	12+log(O/H)	&		RA		&		DEC		&	Exposure	&	Distance	\\
		&			&				&	(J2000)		&	(J2000)		&		(ks)	&		(Mpc)	\\
\hline\noalign{\vspace{1mm}}
DDO68				&	11271	& 7.15	& 09 56 45.7 &	+28 49 35	& 10.00	& 5.90 \\
J081239.5+483645	&	11298	& 7.16	& 08 12 39.5 & +48 36 45 	& 4.78	& 9.04 \\
I Zw 18				&	805		& 7.18	& 09 34 02.4 & +55 14 23 	& 40.77	& 18.20 \\
UGC 772 			&	11281	& 7.24	& 01 13 39.6 & +00 52 31 	& 5.08	& 11.50 \\
J210455.3-003522 	&	11282	& 7.26	& 21 04 55.3 & $-$00 35 22 & 5.01	& 13.70 \\
UGCA 292			&	11295	& 7.27	& 12 38 40.0 & +32 46 01	& 5.01	& 3.50 \\
J141454.1-020823 	&	11293	& 7.32	& 14 14 54.1 & $-$02 08 23	& 16.68	& 24.60 \\
6dF J0405204-364859	&	11292	& 7.34	& 04 05 20.3 & $-$36 49 01	& 5.01	& 11.00 \\
HS 0822+3542		&	11284	& 7.35	& 08 25 55.5 & +35 32 32	& 5.12	& 12.70 \\
SBS 1129+576 		&	11283	& 7.41	& 11 32 02.5 & +57 22 46	& 14.75	& 26.30 \\
KUG 0937+298 		&	11301	& 7.45	& 09 40 12.8 & +29 35 30	& 5.01	& 11.20 \\
J085946.9+392306	&	11299	& 7.45	& 08 59 46.9 & +39 23 06	& 4.78	& 10.90 \\
SBS 0940+544		&	11288	& 7.48	& 09 44 16.6 & +54 11 34	& 16.83	& 22.10 \\
RC2 A1116+51		&	11287	& 7.51	& 11 19 34.3 & +51 30 12	& 11.65	& 20.80 \\
UGC 4483			&	10559	& 7.54	& 08 37 03.0 & +69 46 31	& 3.09	& 3.44 \\
J120122.3+021108	&	11286	& 7.55	& 12 01 22.3 & +02 11 08	& 8.10	& 18.40 \\
KUG 0201-103		&	11297	& 7.56	& 02 04 25.6 & $-$10 09 35	& 13.59	& 22.70 \\
KUG 1013+381		&	11289	& 7.58	& 10 16 24.5 & +37 54 46	& 9.40	& 19.60 \\
SBS 1415+437		&	11291	& 7.59	& 14 17 01.4 & +43 30 05	& 5.11	& 13.70 \\
HS 1442+4250		&	11296	& 7.64	& 14 44 12.8 & +42 37 44	& 5.19	& 8.67 \\
%
%
RC2 A1228+12		&	11290	& 7.64 	& 12 30 48.5 & +12 02 42	& 12.04	& 21.20 \\
SBS 1102+606 		&	11285	& 7.64	& 11 05 53.7 & +60 22 29	& 10.20	& 19.90 \\
%
%
KUG 0942+551 		&	11302	& 7.66	& 09 46 22.8 & +54 52 08	& 16.02	& 24.40 \\
KUG 0743+513 		&	11300	& 7.68	& 07 47 32.1 & +51 11 28	& 5.07	& 8.60 \\
VII Zw 403			&	871		& 7.69	& 11 28 00.4 & +78 59 39	& 10.25	& 3.87 \\
\hline\noalign{\vspace{1mm}}

\end{tabular} 
\end{minipage}
\end{table*}

\section{Analysis}\label{sect:analysis}
\subsection{Chandra Observations}\label{subsect:observations}
All observations were obtained with the back-illuminated ACIS-S3 chip aboard \textit{Chandra} and the galaxies were near the aimpoint on that chip. We reprocessed the level 1 event files using the latest versions of CIAO (4.5.2) and CALDB (4.5.6). X-ray sources were located in the 0.5-8 keV band using the CIAO tool \texttt{wavdetect}. Using the same settings as \cite{mineo}, we created PSF maps using the CIAO tool \texttt{mkpsfmap}, with an enclosed counts fraction (ECF) of 90 percent. The \texttt{wavdetect} tool was run on the $\sqrt{2}$ series of pixel scales from 1 to 8. The significance threshold was set to $10^{-6}$, resulting in less than one false detection per image. We used \texttt{maxiter} = 10, \texttt{iterstop} $= 0.00001$, and \texttt{bkgsigthresh} = 0.0001.

In order to determine if the sources we found using \texttt{wavdetect} were associated with a target galaxy, $D_{25}$ ellipses were constructed using the given dimensions in the HyperLeda\footnote{\url{http://leda.univ-lyon1.fr/}} database. In comparing the $D_{25}$ ellipse for HS 1442+4250 with its optical counterpart, we noticed that the ellipse was not centered and needed to be realigned. The new coordinates were found to be; $\rmn{RA}(\text{J}2000)=14^{\rmn{h}}~44^{\rmn{m}}~12.1\fs2$,
$\rmn{Dec.}~(\text{J}2000)=+42\degr~37\arcmin~37.5\farcs 4$. Using this method, eight sources were found in the sample of 25 galaxies.

Some observations of X-ray binaries have shown them to be removed from the region in which they formed due to kicks during the death of the progenitor to the compact object~\citep{kaaret2004, mapelli2011, rangelov}. These kicks can send the X-ray binaries (that have luminosities $\sim 10^{38}$erg s$^{-1}$ (0.3$-$8.0 keV)) to a distance nearly 200 pc from the star-forming region over the lifetime of the binary. Each of the $D_{25}$ ellipses also have an associated error in their position of 1 arcsec. Taking this into account, we enlarge the dimensions of our ellipses to allow for an extra 200 pc of extent as well as a 1 arcsec positional error. Allowing our ellipses to increase in size, we gain an extra two sources bringing the total to ten.
Additionally, we examined the number of X-ray sources in our sample as a function of distance from their respective BCD. As the size of each ellipse is increased the number of sources it contains grows larger. This is no surprise. However, the background-subtracted number of sources is insensitive to ellipse size (remaining at $\sim$ 10 sources) once the dimensions of the ellipses are increased by 20 per cent of their original size. Increasing the dimensions of the ellipses as previously described corresponds to a 25 per cent increase on average. All ten sources are less than 1.5 arcmin off-axis.

The number of background sources, $N_\text{bkg}$, was estimated using the $\log N - \log S$ curves of \citet{georgakakis}, which combines both shallow and deep X-ray observations. The flux limits used for these estimates were calculated assuming a minimum of 9 counts for source detection (as discussed in the following section). We find that upon increasing the size of each ellipse, the total number of expected background sources is $N_\text{bkg} = 0.76$, an increase of $<0.2$ sources. Thus the additional two sources are likely associated with their respective target galaxies and we continue the analysis, keeping these additional sources while also taking into account the expanded $D_{25}$ when calculating SFR estimates.

\subsection{Completeness}\label{subsect:completeness}
From \citet{zezas}, the probability of detecting a source given the background counts per pixel and the source counts is given by,
\begin{equation}\label{eqn:zezas}
A(C) = 1.0 - \lambda_0 C^{-\lambda_1} e^{-\lambda_2 C}.
\end{equation}
Here $C$ is the total number of counts from the source and $\lambda_0,\ \lambda_1,\ \lambda_2$ are parameters that depend on the background counts per pixel. The probability relation was found using sources less than 1.5 arcmin off-axis. All 10 of our sources meet this criterion. For our sample, the background is less than 0.025 counts per pixel for all observations. This is the lowest background \cite{zezas} used in their data sets. Thus, the completeness estimate is a conservative since we are assuming a higher background for this calculation. Their fit for this background level gives us parameters such that,
\begin{equation}\label{eqn:zezas_fit}
A(C) = 1.0 - 11.12 C^{-0.83} e^{-0.43 C}.
\end{equation}

In order to detect sources with at least a 95 percent probability, there must be a minimum of 9 source counts in the $0.5 - 8$ keV band. From this threshold, we calculated a flux using the \texttt{WebPIMMS}\footnote{\url{http://heasarc.gsfc.nasa.gov/Tools/w3pimms.html}} tool. The input parameters for this calculation are the count rate (9 counts/(exposure time)), the neutral hydrogen column density (found using \texttt{Colden}\footnote{\url{http://cxc.harvard.edu/toolkit/colden.jsp}}), the photon index ($\Gamma = 1.7$) and the energy range ($0.5 - 8$ keV). This minimum flux was then converted to a luminosity using the distance to the target given in Table~\ref{PropTable}. This luminosity is the assumed completeness luminosity. All sources associated with our sample of BCDs meet the required 9 count minimum.

\begin{table*}
\centering
\begin{minipage}{155mm}
\caption{X-Ray Source Photometry}\label{AnalysisTableN}
\begin{tabular}{l c c l r c c c c}
\hline
\hline\noalign{\vspace{1mm}}
Name	&	${n_H}^a$	&		\multicolumn{2}{c}{${D_{25}}^b$}	&		angle$^b$		&	${N(>L_\text{min})}^c$	&	${N_\text{bkg}}^d$	& ${L_\text{min}}^e$	& ${L_\text{source}}^f$ \\
		&	$(10^{20} \text{cm}^{-2})$		&	\multicolumn{2}{c}{(arcsec)}	&	(deg)	&	& & $\left(10^{38} \text{erg s}^{-1}\right)$	& $\left(10^{38} \text{erg s}^{-1}\right)$	\\
\hline\noalign{\vspace{1mm}}
I Zw 18                 & 1.99  &       22.7	&  17.9	&  145.0 &       1     	& 0.1002        & 0.737	& 38.10 \\
VII Zw 403              & 3.91  &       71.0	&  48.3	&  6.0   &       2     	& 0.0774        & 0.137	& 3.982, 0.151 \\
UGC 4483                & 3.23  &       93.3	&  63.0	&  169.3 &       0     	& 0.0379        & 0.355	& $-$\\
DDO 68                  & 1.97  &       135.7	&  56.5	&  13.4  &       2     	& 0.1597        & 0.316	& 0.489, 0.620\\
UGC 772                 & 3.10  &       81.3	&  63.9	&  0.0   &       0     	& 0.0417        & 2.404	& $-$\\
J210455.31-003522.2     & 5.66  &       24.2	&  18.5	&  177.9 &       0     	& 0.0036        & 3.609	&  $-$\\
SBS 1129+576            & 0.87  &       47.6	&  19.2	&  162.0 &       1    	& 0.0372        & 4.173	& 10.14 \\
HS 0822+3542            & 4.82  &       25.8	&  16.8	&  141.4 &       0    	& 0.0036        & 2.995	& $-$\\
SBS 1102+606            & 0.59  &       41.5	&  24.7	&  104.8 &       0		& 0.0215        & 3.438	& $-$\\
J120122.3+021108.5      & 1.88  &       33.3	&  20.6	&  129.2 &       0		& 0.0100        & 3.784	& $-$\\
RC2 A1116+51            & 1.19  &       24.1	&  18.8	&  90.9  &       1		& 0.0122        & 3.322	& 32.07\\
SBS 0940+544            & 1.34  &       39.5	&  22.2	&  157.2 &       1		& 0.0458        & 2.820	& 9.80\\
KUG 1013+381            & 1.41  &       26.5	&  24.3	&  40.7  &       0     	& 0.0119        & 3.670	& $-$\\
RC2 A1228+12            & 2.47  &       24.9	&  20.3	&  78.8  &       1     	& 0.0147        & 3.414	& 3.414\\
SBS 1415+437            & 1.21  &       28.8	&  18.2	&  58.0  &       0     	& 0.0042        & 3.288	& $-$\\
6dF J0405204-364859     & 0.88  &       37.6	&  32.3	&  16.5  &       0     	& 0.0094        & 2.150	& $-$\\
J141454.13-020822.9     & 4.17  &       24.8	&  18.5	&  28.4  &       0     	& 0.0249        & 3.413	& $-$\\
UGCA 292                & 1.34  &       84.2	&  66.1	&  0.0   &       0     	& 0.0433        & 0.219	& $-$\\
HS 1442+4250            & 1.53  &       63.8	&  23.2	&  63.2  &       1     	& 0.0119        & 1.303	& 2.164\\
KUG 0201-103            & 2.08  &       31.9	&  16.1	&  101.3 &       0     	& 0.0185        & 3.445	& $-$\\
J081239.52+483645.3     & 4.58  &       36.7	&  25.2	&  73.7  &       0     	& 0.0071        & 1.619	& $-$\\
J085946.92+392305.6     & 2.44  &       35.2	&  27.7	&  85.0  &       0     	& 0.0074        & 2.272	& $-$\\
KUG 0743+513            & 5.17  &       62.7	&  36.6	&  124.2 &       0     	& 0.0188        & 1.395	& $-$\\
KUG 0937+298            & 1.81  &       46.4	&  24.1	&  92.2  &       0     	& 0.0087        & 2.265	& $-$\\
KUG 0942+551            & 1.23  &       27.7	&  20.1	&  68.2  &       0     	& 0.0266        & 3.327	& $-$\\
\hline\noalign{\vspace{1mm}}
\end{tabular}
\\
\textbf{Notes.}\\
$^a$ Column densities were found using the \texttt{Colden}\footnotemark[4] tool.\\
$^b$ Properties of the $D_{25}$ ellipses were taken from the HyperLeda\footnotemark[2] database and modified as described in Section~\ref{subsect:observations}.\\
$^c$ Number of observed sources within the $D_{25}$ ellipse with luminosity greater than $L_\text{min}$.\\
$^d$ Number of expected background sources within the $D_{25}$ ellipse, as determined by $\log N - \log S$ curves of \citet{georgakakis}.\\
$^e$ Luminosities limits calculated for the $0.5-8.0$ keV energy band.\\
$^f$ Luminosity of detected source(s) in the $0.5-8.0$ keV energy band.
\end{minipage}

\end{table*}

\subsection{Star Formation Rates}\label{subsect:headsfr}
Star formation rates were determined using two methods. The first method relates the FUV luminosity of the dwarf galaxy to SFR, as found by \citet{hunter} and allows us to calculate a SFR for each galaxy in the sample. The second method was employed in order to be consistent with \citet{mineo} and accounts for UV light escaping the galaxy as well as IR emission due to heating of dust by young stars. However, this method cannot be applied across the whole sample due to lack of observations in the IR. This problem is discussed further in Section~\ref{subsect:sfr2}.
Observations made by \textit{Galaxy Evolution Explorer} (\textit{GALEX}) were used for the UV data and observations by \textit{Spitzer} were used in determining the IR contribution.
\subsubsection{SFR via \textit{GALEX} Observations in the FUV}\label{subsect:sfr}
Images corresponding to each galaxy in the sample were found in the \textit{GALEX} archive. In order to calculate the SFR for each galaxy, we follow the procedure of \citet{prestwich}. We first needed to extract the count rate for each galaxy. The pixel values for the image files (*int.fits) report photons/pixel/second corrected for relative response. By using the previously constructed $D_{25}$ ellipses as source apertures, we extracted background-subtracted FUV count rates for the galaxies. For background apertures, we used annular ellipses centered on the source galaxy with dimensions scaled to give $8\times$ the area of the source aperture. Using the CIAO \texttt{dmstat} tool, we obtained the background-subtracted count rates for each galaxy in the FUV band.

For \textit{GALEX}{\footnote{\href{http://galexgi.gsfc.nasa.gov/docs/galex/FAQ/counts_background.html}{\nolinkurl{http://galexgi.gsfc.nasa.gov/}}}}, the specific flux is proportional to the count rate $r$,
\[
f_\text{FUV} = (1.4\times 10^{-15})\times r \text{ [erg cm$^{-2}$ s$^{-1}$ \AA$^{-1}$]}.
\]
Following \citet{prestwich}, we convert units from \AA\ to Hz in order to have the correct units for luminosity in Eq.~(\ref{SFR_eq}).
\[
f_\text{FUV} = (1.09\times 10^{-27})\times r \text{ [erg cm$^{-2}$ s$^{-1}$ Hz$^{-1}$]}
\]
From these fluxes, we calculate a luminosity $L_\text{FUV}$ that is corrected for extinction due to Galactic dust. This is done using the method given in \citet{cardelli}. The reddening magnitudes found in Table~\ref{AnalysisTableSFR}, were computed using the Infrared Science Archive (IRSA) tool DUST~\footnote{\url{http://irsa.ipac.caltech.edu/applications/DUST/}}. Then, using the relation found in \citet{hunter}, we relate the specific luminosity to the SFR,
\begin{equation}\label{SFR_eq}
\text{SFR}_\text{FUV} (M_\odot\ \text{yr}^{-1}) = 1.27\times 10^{-28}\ L_\text{FUV}\ (\text{erg s$^{-1}$ Hz$^{-1}$}).
\end{equation}

\subsubsection{SFR via NUV and IR Luminosities}\label{subsect:sfr2}
Observations in the NUV band were found in the \textit{GALEX} archive for each galaxy in the sample. However, only 13 out of 25 galaxies in our sample had been observed by \textit{Spitzer}. 
This smaller sample of galaxies contains 9 of the 10 observed X-ray sources and nearly 70 per cent of the total star-formation rate as determined by the previous method. 
In order to compare our results with \citet{mineo}, we determine SFRs for this smaller sample of galaxies, which dominates in its overall contribution to SFR, and find a relation that allows us to estimate the SFR for the other 12 galaxies.

For the \textit{GALEX} NUV images, we use the same approach that we used for the FUV band to calculate a flux,
\[
f_\text{NUV} = (2.06\times 10^{-16})\times r \text{ [erg cm$^{-2}$ s$^{-1}$ \AA$^{-1}$]},
\]
where $r$ is still the background-subtracted count rate. Using this flux we calculate a monochromatic luminosity $($erg s$^{-1}$\AA$^{-1})$ at 2312 \AA\ for NUV which is not corrected for extinction.

We used multiband imaging photometer for \textit{Spitzer} (MIPS) 70-$\mu$m Large Field images to find the IR component of the SFR. Following \citet{mineo}, we used the post Basic Calibrated Data (pBCD) products found in the the \textit{Spitzer} archive\footnote{\url{http://irsa.ipac.caltech.edu/applications/Spitzer/}}\label{foot:spitzer}. We used the same process to extract background-subtracted counts as we did for both UV data. However, for some images the background aperature's size needed to be reduced or rotated due to the sources proximity to the image's edge and/or other bright sources. The images are calibrated in units of MJy sr$^{-1}$ and were converted to Jy using the following conversion factor\footnotemark[7]:
\[
C_{70\mu \text{m}} = 3.76\times 10^{-4}.
\]
The flux for each galaxy was then converted to a spectral luminosity $($erg s$^{-1}$ Hz$^{-1})$ which was then used to estimate the total IR luminosity\footnotemark[7] $(8-1000\mu $m$)$ using,
\[
L_\text{IR}(L_{\odot}) = 7.9\times (\nu L_{\nu})^{0.94}_{70\mu \text{m}},
\] 
where $L_{\odot} = 3.839\times 10^{33}$ erg s$^{-1}$.

From our luminosity measurements we calculated the total SFR,
\[
\text{SFR}_\text{tot} = \text{SFR}_{\text{NUV},0} + (1-\eta)\text{SFR}_\text{IR} ,
\]
where SFR$_{\text{NUV},0}$ and SFR$_\text{IR}$ are obtained from,
\begin{align*}
\text{SFR}_\text{IR} (M_\odot \text{yr}^{-1}) 		&= 4.6\times 10^{-44} L_\text{IR} (\text{erg s}^{-1})\\
\text{SFR}_{\text{NUV},0} (M_\odot \text{yr}^{-1})	&= 1.2\times 10^{-43} L_\text{NUV,obs} (\text{erg s}^{-1}).
\end{align*}
In order to account for IR emission due to an old stellar population in normal star-forming galaxies, a correction factor $\eta$ is used. For starburst galaxies, $\eta \approx 0$.
$L_\text{NUV,obs}$ is the observed NUV (2312 \AA) luminosity, which is uncorrected for dust attenuation, and $L_\text{IR}$ is the 8$-$1000 $\mu$m luminosity.
Results for both methods can be found in Table~\ref{AnalysisTableSFR}.\\

\subsubsection{Comparison of Methods Used in Determining SFR}
\begin{figure}
\centering
\includegraphics[width=0.5\textwidth]{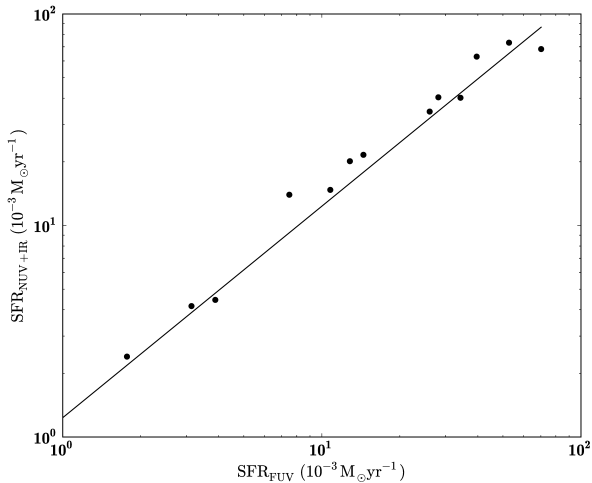}
\caption{Comparison of the two methods used for determining star formation rate. The SFR relation used by \citet{mineo} determines a rate that is a factor of $1.23\pm 0.11$ larger than using the method employed by \citet{prestwich}.}\label{fig:sfr_comparison}
\end{figure}

In Figure~\ref{fig:sfr_comparison}, we compare the two methods used to calculate SFR for the sample of 13 galaxies that were observed in all bands. We find a linear correlation between the two methods that shows the method used by \citet{mineo} produces a SFR that is a factor of $1.23\pm 0.11$ larger than the FUV method of \citet{hunter}. This is consistent with the 26 per cent increase that \citet{hunter} find when
comparing their method to that of \citet{iglesias2006} that was used by \citet{mineo}. \citet{iglesias2006} designed their method for normal, solar metallicities galaxies, whereas \citet{hunter} constructed their relation specifically for dwarf galaxies with sub-solar metallicities. Thus, the \citet{hunter} relation should be more appropriate for our sample. However, we used the relation of \citet{mineo} to allow a direct comparison with their paper. We did this by using the correlation to estimate the SFRs for the remaining 12 galaxies that were not observed in the IR.
\begin{table*}\scriptsize
\centering
\begin{minipage}{155mm}
\caption{UV, IR Measurements and SFRs.}\label{AnalysisTableSFR}
\begin{tabular}{l c c c c c c c}
\hline
\hline\noalign{\vspace{1mm}}
Name	&	$E(B-V)^a$ & Count Rate$_\text{FUV}^b$ & $L_\text{FUV}^c$ &	$\text{SFR}_{\text{FUV}}^d$	& $\text{SFR}_{\text{NUV+IR}}^e$	&	Count Rate$_\text{NUV}^f$	&	Flux$_\text{IR}^g$ \\
		& (mag)	& (cps) & $\left(10^{26} \text{erg s}^{-1} \text{Hz}^{-1}\right)$	&	$\left(10^{-3} M_{\odot}\ \text{yr}^{-1}\right)$	& $\left(10^{-3} M_{\odot}\ \text{yr}^{-1}\right)$	&	(cps)	&	(mJy) \\
\hline\noalign{\vspace{1mm}}
I Zw 18                 & 0.032 &       10.766  &       5.528	&70.20	& 68.27	&	28.64	& 14.01\\
VII Zw 403              & 0.037 &       24.061  &       0.590 &7.49 	& 13.98	&	76.31	& 586.96\\
UGC 4483                & 0.033 &       12.575  &       0.247	&3.14 	& 4.16	&	40.24	& 95.74\\
DDO 68                  & 0.018 &       25.189  &       1.011 &12.83	& 20.14	&	61.97	& 218.56\\
UGC 772                 & 0.028 &       4.951   &       0.751 &9.54	& $-$ 	&	12.32	& $-$\\
J210455.31-003522.2     & 0.066 &       1.241   &       0.468 &5.94	& $-$ 	&	3.58	& $-$\\
SBS 1129+576            & 0.013 &       3.576   &       3.112 &39.59	& 62.88	&	11.25	& 20.51\\
HS 0822+3542            & 0.047 &       1.076   &       0.306 &3.88	& 4.45	&	3.63	& 3.31\\
SBS 1102+606            & 0.006 &       4.344   &       2.053 &26.08	& 34.57	&	11.28	& 14.05\\
J120122.3+021108.5      & 0.024 &       1.209   &       0.519 &6.59 	& $-$ 	&	3.33	& $-$\\
RC2 A1116+51            & 0.015 &       4.579   &       2.699 &34.28	& 40.18	&	13.31	& 2.32\\
SBS 0940+544            & 0.013 &       1.631   &       1.140 &14.47	& 21.59	&	5.61	& 7.94\\
KUG 1013+381            & 0.015 &       7.945   &       4.155 &52.77	& 73.19	&	24.25	& 36.13\\
RC2 A1228+12            & 0.027 &       2.920   &       1.955 &24.83	& $-$ 	&	9.09	& $-$\\
SBS 1415+437            & 0.009 &       9.962   &       2.217 &28.16	& 40.38	&	25.95	& 54.10\\
6dF J0405204-364859     & 0.006 &       5.592   &       0.857 &10.88	& $-$ 	&	17.01	& $-$\\
J141454.13-020822.9     & 0.058 &       0.670   &       0.733 &9.31	& $-$ 	&	2.20	& $-$\\
UGCA 292                & 0.016 &       8.797   &       0.139 &1.77 	& 2.40	&	23.28	& 44.50\\
HS 1442+4250            & 0.013 &       9.156   &       0.849 &10.78	& 14.74	&	26.52	& 18.67\\
KUG 0201-103            & 0.021 &       1.383   &       0.952 &12.08	& $-$ 	&	4.47	& $-$\\
J081239.52+483645.3     & 0.051 &       0.897   &       0.130 &1.65 	& $-$ 	&	2.99	& $-$\\
J085946.92+392305.6     & 0.026 &       0.743   &       0.131 &1.67 	& $-$ 	&	3.32	& $-$\\
KUG 0743+513            & 0.070 &       13.290  &       2.041 &25.92	& $-$ 	&	38.53	& $-$\\
KUG 0937+298            & 0.018 &       5.649   &       0.976 &12.40	& $-$ 	&	17.00	& $-$\\
KUG 0942+551            & 0.012 &       1.555   &       1.223 &15.53	& $-$ 	&	5.14	& $-$\\

\hline\noalign{\vspace{1mm}}
\end{tabular}
\footnotesize
\\
\textbf{Notes.}\\
$^a$ Extinctions were computed using the Infrared Science Archive (IRSA) tool DUST.\\
$^b$ Count rate in the far ultraviolet band (FUV) of GALEX observations.\\
$^c$ Luminosity in the FUV band, $1350-1750$\AA . \\
$^d$ Star-formation rate as determined by the \citet{hunter} method.\\
$^e$ Star-formation rate as determined by the \citet{iglesias2006}.\\
$^f$ Count rate in the near ultraviolet band (NUV) of GALEX, $1750-2800$\AA .\\
$^g$ Infrared flux in mJy from Spitzer in the $8-1000\ \mu$m range.
\end{minipage}
\end{table*}
\section{Bayesian Approach to Finding the XLF}\label{sect:bayes}

From \citet{grimm}, the luminosity function for normal metallicity galaxies is of the form,
\begin{equation}\label{lumFunct}
\frac{dN}{dL_{38}}=q\, s L_{38}^{-\alpha}\ \ \text{for}\ \ L\leq L_{\text{cut}}
\end{equation}
where $L_{38}=L_X/(10^{38}\, \text{erg s}^{-1})$, $q$ is a normalisation constant, $\alpha$ is the power law index, $s$ is the star formation rate $(M_{\odot}\ \text{yr}^{-1})$, and $L_\text{cut}$ is the cut-off luminosity for the XLF.
For data that are complete down to some luminosity, $L_\text{min}$, the expected number of sources in the range $\left[L_\text{min},L_\text{cut}\right]$ $(10^{38}$ erg s$^{-1})$ is found by integrating Eq.~(\ref{lumFunct}),
\begin{equation}\label{expNum}
N(>L_\text{min}) = q \frac{s}{1-\alpha}\left(L_\text{cut}^{1-\alpha} - L_\text{min}^{1-\alpha}\right).
\end{equation}
Given that Eq.~(\ref{expNum}) describes an expected number of sources over a certain luminosity range, the number of sources that are actually observed should follow a Poisson distribution. That is,
\begin{equation}\label{poisson}
P(x;N(>L)) = \frac{e^{-N}N^x}{x!}\ .
\end{equation}

It should be mentioned that a Poisson distribution is being assumed as an approximation to what is actually a Binomial distribution, as has been pointed out by \citet{kelly}. Since the total number of X-ray sources at \textbf{all} luminosities in a galaxy is finite, and the number seen in a certain luminosity range is some fraction of this total, the appropriate distribution to use would be a Binomial distribution. However, if the number of observed sources is much smaller than the total number of sources, the Binomial distribution may be approximated by the Poisson distribution. 

In order to calculate a probability distribution for our normalisation parameter $q$ from the SFR and luminosity limits, we apply Bayes' theorem.
Bayes' theorem states that,
\begin{equation}\label{bayesthm}
f'(q|D) = \frac{P(D|q)f(q)}{\int P(D|q)f(q)dq},
\end{equation}
where $D$ represents some observed data, $q$ is the desired parameter, the function $f'$ is the posterior probability distribution, $P$ is the likelihood function, and $f$ is the prior probability distribution. In our case, we calculate a posterior probability distribution for the XLF normalisation $q$. We do this by using the Poisson distribution in Eq.~(\ref{poisson}) as our likelihood function $P(D|q)$, where the expected number of sources $N$ comes from Eq.~(\ref{expNum}). The data $D$ will be composed of our observed number of X-ray sources, SFR, and luminosity limits. The prior, $f(q)$, provides an initial guess for the probability distribution of $q$. The Bayesian method requires calculating $f'(q|D)$, using Eq.~(\ref{bayesthm}), iteratively for each galaxy. That is, $f'(q|D)$ for one galaxy becomes $f(q)$ for the next.

In doing the calculation associated with Eq.~(\ref{bayesthm}), one then finds that the posterior has the form of the gamma distribution,
\begin{equation}\label{gammaDist}
\text{GAMMA}(q;X,B) = \frac{B^{X} q^X e^{-qB}}{(X-1)!},
\end{equation}
where $X = \sum x_i$, $x_i$ is the number of observed sources in the $i^\text{th}$ galaxy, and $B = \sum \frac{s_i}{1-\alpha}\left(L_\text{cut}^{1-\alpha} - L_\text{min,\ i}^{1-\alpha}\right)$.

As an initial step, we assume a prior of $\text{GAMMA}(q;0,0)$. We call this prior the \textit{know-nothing}\footnote{\url{http://aleph0.clarku.edu/~djoyce/ma218/bayes3.pdf}} Gamma distribution. The assumption of this prior yields the same results as a frequentist approach.

The Gamma distribution is also known as the \textit{conjugate prior} to the Poisson distribution. That is, if the likelihood is modelled by a Poisson distribution, the posterior will be a Gamma distribution. This allows us to skip the task of iteratively calculating a posterior and simply jump straight to Eq.~(\ref{gammaDist}). Some of the important properties of the Gamma distribution are given in Table~\ref{GammaDistProps}.

Our calculation should weight the luminosity function by the completeness function, given by Eq.~\ref{eqn:zezas_fit}, in order to account for the probability that a source exists but is not detected. In Eq.~\ref{eqn:zezas_fit}, the number of counts, $C$, depends linearly on the luminosity, so we replace it with $C = b L$, where $b$ contains the distance and conversion factors.
Thus the weighted luminosity function should be written as,
\begin{equation}
\frac{dN}{dL_{38}} = q s L_{38}^{-\alpha}\left(1.0-11.12 (b L)^{-0.83} e^{-0.43 b L}\right).
\end{equation} 
Integrating this equation from $L_\text{min}$, which is equivalent to $C=9$ by definition, to $L_\text{cut} = 110\ (\times 10^{38}$ erg s$^{-1})$, one finds the total number of expected X-ray sources weighted by completeness.
If one replaces Eq.~\ref{expNum} with this result and carries out the calculation as before, one finds that the final result changes the normalization $q$ by 0.4 per cent. The error on the final value of $q$ is about 33 per cent (see below), thus we are justified in neglecting the source detection probability.

\begin{table}
\centering
\caption{Properties of the Gamma Distribution}\label{GammaDistProps}
\begin{tabular}{l l}
\hline
\hline\noalign{\vspace{1mm}}
Form			& $\text{GAMMA}(q;X,B) = \left(B^{X} q^{X-1} e^{-qB}\right)/(X-1)!$\\
Mean			& $X/B$\\
Mode 			& $ (X-1)/B\ \ \text{for  }X>1$\\
Median 			& $q_0$, where $0.5 = \int_0^{q_0} \text{GAMMA}(q;X,B)dq$\\
Variance 		& $ X/B^2$\\

\hline\noalign{\vspace{1mm}}
\end{tabular}

\end{table}
\section{Results}\label{sect:results}
In this paper, the galaxies observed are metal poor or extremely metal poor (XMPG). The luminosity function found by \citet{grimm} is for normal metallicity galaxies. We applied the same functional form for the luminosity function given by \citet{grimm} to these metal poor galaxies. Following \citet{mineo}, we assumed $\alpha = 1.58$, $L_{\text{cut}} = 110 \times 10^{38} \text{erg s}^{-1}$, and we used the SFRs determined by their method for galaxies that had UV and IR observations. For the 12 galaxies that were not observed in the IR, we scaled the SFRs determined from the FUV by a factor of 1.23, as discussed at the end of Section~\ref{subsect:headsfr}.

\begin{figure}
\centering
\includegraphics[width=0.5\textwidth]{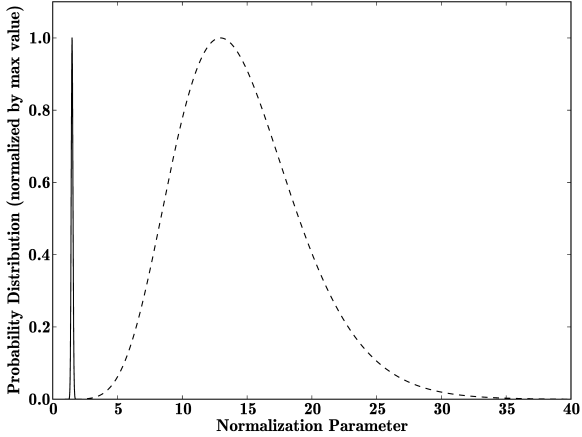}
\caption{Normal galaxies (solid line; Mineo et al.) $q = 1.49\pm 0.07\ (M_\odot^{-1}\, \text{yr})$, low metallicity BCDs (dashed line; Brorby) $q = 14.5\pm 4.8\ (M_\odot^{-1}\, \text{yr})$. The difference in appearance (and lack of overlap) in the probability distributions of the normalisation parameter is evident for the two different populations of galaxies. By calculating their overlap, we may exclude the hypothesis that the XLF normalisation parameters agree with $1-5\times 10^{-6}$ confidence. The probability distributions are normalized with respect to their peak values (as opposed to area) in order to enhance visual comparison.}\label{ProbDist_Brorby_Grimm_prior1}
\end{figure}

From Figure~\ref{ProbDist_Brorby_Grimm_prior1}, one can see the probability distribution for our sample of BCDs. In the \citet{mineo} paper, a value of $q = 1.49\pm 0.07\ (M_\odot^{-1}\, \text{yr})$ is given as the normalisation parameter in Eq.~(\ref{lumFunct}). From our analysis, we calculate an XLF normalisation of $q = 14.5\pm 4.8\ (M_\odot^{-1}\, \text{yr})$. Thus, by calculating the overlap between the two distributions we find that we can exclude the hypothesis that normal and BCD galaxies have the same normalisation parameter, $q$, with probability of $1-5\times 10^{-6}$.

As an additional check to these statistical calculations, we use a frequentist approach to find the likelihood of detecting 10 or more X-ray sources given the model of \citet{mineo}. From our SFRs, luminosity limits, expected background sources, and using the XLF normalisation of 1.49, we expect to detect 1.75 sources. Assuming Poisson statistics for the number of sources detected with a mean of 1.75, detecting 10 or more sources has a probability of $1.5\times 10^{-5}$.

Using the SFRs we calculated from the FUV measurements, we find a normalisation of $q = 17.9\pm 5.9\ (M_\odot^{-1}\, \text{yr})$.

\subsection{Uncertainty in SFR}\label{subsect:uncertainty}
Our results do not take measurement uncertainty of SFR into account. All quoted uncertainties are statistical in nature. To this end, we now devise a method in which we may account for the uncertainty in SFR measurements. 

When calculating star formation rates, a typical observational uncertainty of $\sim 30$ percent is quoted. This arises from complexity in the physical processes of the system which produce scatter in the data around the simplified models used to describe SFR as functions of luminosity in bands such as UV or IR~\citep{kennicutt}.

\begin{figure}
\centering
\includegraphics[width=0.5\textwidth]{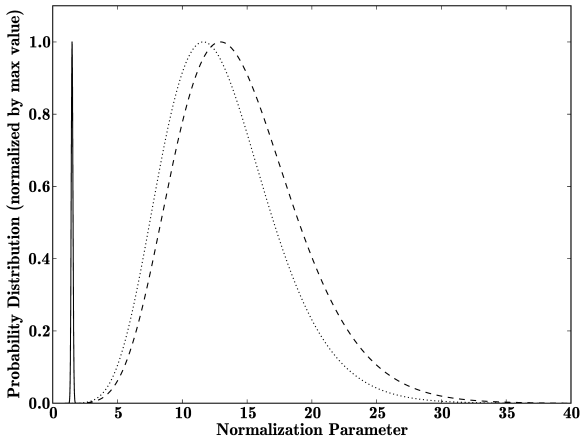}
\caption{Normal galaxies (solid line; Mineo et al.) $q = 1.49\pm 0.07\ (M_\odot^{-1}\, \text{yr})$, BCDs (dashed line; Brorby) $q = 14.5\pm 4.8\ (M_\odot^{-1}\, \text{yr})$, Lower limit for BCDs (dotted line; Brorby) $q = 13.0\pm 4.3\ (M_\odot^{-1}\, \text{yr})$. Same as Figure~\ref{ProbDist_Brorby_Grimm_prior1}, but with a lower limit PDF that was calculated based on SFR uncertainty. The \textit{lower limit} PDF excludes the \citet{mineo} value for the XLF normalisation $q$ with a confidence of $1-10^{-5}$}\label{pdf_lower}
\end{figure}

Assuming the errors in SFR are Gaussian about our calculated values, we generated $10^5$ samples of galaxies and calculated the mean and standard deviation. The mean matched the original (14.5, as expected) and the standard deviation was 1.5. Using this information, we calculated the probability distribution function (PDF) for an XLF normalisation with $q = 14.5-1.5 = 13.0$. This new \textit{lower limit} PDF excludes the \citet{mineo} value with a confidence of $1-10^{-5}$ (see Figure~\ref{pdf_lower}). This gives us further confidence in the significance of our result.

\begin{figure}
\centering
\includegraphics[width=0.4\textwidth]{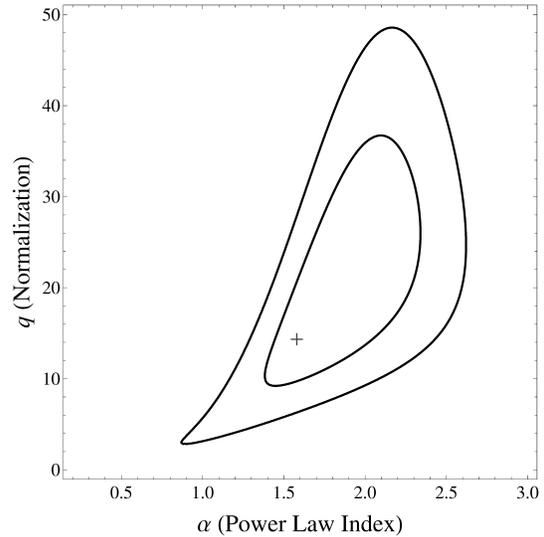}
\caption{Contours correspond to 68 per cent and 95 per cent confidence regions. The most probable values for the XLF normalisation and power law index are $q = 21.2^{+12.2}_{-8.8}\ (M_\odot^{-1}\, \text{yr})$ and $\alpha = 1.89^{+0.41}_{-0.30}$, respectively. The cross represents the location of our reported value of $q = 14.5\pm 4.8\ (M_\odot^{-1}\, \text{yr})$ and our assumed values of $\alpha = 1.58$.}\label{fig:contour}
\end{figure}
\subsection{Uncertainty in Power Law Index $\alpha$}
The Bayesian calculation may be extended to two or more parameters in order to determine multivariate probability distribution functions. Using the methods described in Section~\ref{sect:bayes}, we allowed both the XLF normalisation $q$ and the power law index $\alpha$ to be simultaneously described by a two-variable probability distribution function. The 68 per cent and 95 per cent confidence regions are shown as contours in Figure~\ref{fig:contour}. Finding the point of maximum probability for the PDF produces simultaneous estimates for both $q$ and $\alpha$. In our case, $q = 21.2^{+12.2}_{-8.8}\ (M_\odot^{-1}\, \text{yr})$ and $\alpha = 1.89^{+0.41}_{-0.30}$ are the most probable values for the XLF normalisation and power law index, respectively. These values are consistent with our assumed, fixed value of $\alpha = 1.58$ and the corresponding XLF normalisation of $q = 14.5\pm 4.8\ (M_\odot^{-1}\, \text{yr})$, which are represented as a cross in Figure~\ref{fig:contour}.
\section{Summary and Conclusions}\label{sect:conclusions}
Being nearby analogues to low-metallicity galaxies in the early universe, evidence for enhanced HMXB populations in BCDs could suggest HMXBs played a major role in the Epoch of Reionization. The analysis in this paper shows that the population of these X-ray binaries in BCDs is indeed significantly larger than the expected value calculated from the XLF for near-solar metallicity galaxies, thus implying a heightened HMXB population in the early universe.

Our analysis shows that for our sample of known BCDs the XLF normalisation is larger than that of near-solar metallicity galaxies by a factor of $9.7\pm 3.2$. Specifically, we find a normalisation of $q = 14.5\pm 4.8\ (M_\odot^{-1}\, \text{yr})$ as compared to the value of \citet{mineo}, $q = 1.49\pm 0.07\ (M_\odot^{-1}\, \text{yr})$, which is consistent with previous studies of near-solar metallicity galaxies. Simultaneous determination of the XLF normalisation and power law index yields $q = 21.2^{+12.2}_{-8.8}\ (M_\odot^{-1}\, \text{yr})$ and $\alpha = 1.89^{+0.41}_{-0.30}$. Thus, we find that the power law index for our sample of low-metallicity BCDs is consistent with that of near-solar metallicity galaxies. This suggests that the enhanced production of HMXBs relative to SFR may be modelled by the same XLF as normal galaxies, except with an enhanced normalisation parameter.

Previous studies have also found enhanced production of X-ray binaries relative to SFR in low metallicity galaxies \citep[e.g.][]{mapelli2010,kaaret,prestwich,basu-zych2013b}. The low metallicity environments in which these objects were formed is thought to be similar to the first small galaxies of the early universe \citep{kunth}. These findings suggest X-ray production was enhanced relative to SFR in the early universe, thus making X-rays an important and previously overlooked contributor to heating during the Epoch of Reionisation~\citep{mirabel}.

Using observational data from $z\leq 4$ surveys, \cite{basu-zych2013a} showed that the $L_X$/SFR relation evolved with redshift, which they ascribed to changes in the HMXB metallicities. Building upon this work, \cite{fragos2013a} used previous large scale simulations to synthesize populations of X-ray binaries from the time galaxies first were forming up to the present time. The same observational data used by \cite{basu-zych2013a} were used to constrain their models from which they found that at $z\geq 2.5$ HMXBs began to dominate the X-ray emission over LMXBs and AGN, increasing by about a factor of 5 in their X-ray luminosity relative to SFR ($L_X$/SFR).

Most simulations to date have assumed a power law spectrum in the X-rays for these HMXBs~\citep[e.g.,][]{mesinger2013,mcquinn2012}. However, \cite{kaaret2014} has recently shown that the total spectrum of these early universe populations are likely dominated by HMXBs with curved spectra in high luminosity states, affecting and weakening constraints on emission from HMXBs in the soft X-ray background. Curved spectra would also modify the K-correction in such a way that is consistent with a large enhancement of $L_X$/SFR at redshifts of $z=6-7$ when applied to deep X-ray surveys.

All of these results may be used to provide better estimates of the X-ray contribution to the heating of the early universe. The conditions associated with this heating affect the ionisation morphology during the Epoch of Reionisation via thermal feedback, and also affect the time at which reionisation was completed.

\label{lastpage}

\end{document}